%% file: paper.tex
\documentclass{llncs}

\usepackage{url}
\usepackage{breakurl} 
\usepackage[breaklinks]{hyperref}
\usepackage[table]{xcolor} 
\newcolumntype{?}{!{\vrule width 0.75pt}}
\usepackage{makecell}
\usepackage[T1]{fontenc} 
\usepackage[utf8]{inputenc}
\usepackage{times}

\usepackage{textcomp}

\usepackage{balance}
\makeatletter
\renewcommand\large{\@setfontsize\large\@xiipt{15\p@}}
\makeatother
\usepackage[tracking=true]{microtype}
\usepackage{booktabs}
\usepackage{tikz}
\usetikzlibrary{arrows,positioning}
\usepackage{cite}
\usepackage{multirow}
\usepackage{rotating}

\input{quic-tlds.tex}

\input{quic-isp-shares.tex}

\input{quic-Mobileisp-shares.tex}

\input{quic-mawi-shares.tex}

\input{quic-asinfos.tex}

\input{quic-latesthttpgrep.tex}

\input{quic-certificate-data.tex}

\input{quic-DeCIX-shares.tex}

\input{content-valuesredef}

\usepackage{tikz-qtree}
\usetikzlibrary{shadows,trees}

\newcommand{\afblock}[1]{\noindent{\textbf{#1 }}}
\newcommand{\takeaway}[1]{\noindent{\textbf{Takeaway.}} \textit{#1}}
\usepackage{makeidx}  

\newcommand\copyrighttext{%
  \footnotesize The final publications is available at Springer via \url{http://dx.doi.org/10.1007/978-3-319-76481-8_19}}
\newcommand\copyrightnotice{%
\begin{tikzpicture}[remember picture,overlay]
\node[anchor=south,yshift=10pt] at (current page.south) {\fbox{\parbox{\dimexpr\textwidth-\fboxsep-\fboxrule\relax}{\copyrighttext}}};
\end{tikzpicture}%
}

\begin{document}
\frontmatter          
\mainmatter              
\title{A First Look at QUIC in the Wild}
\titlerunning{A First Look at QUIC in the Wild}  
\author{Jan R\"uth\inst{1} \and Ingmar Poese\inst{2} \and Christoph Dietzel\inst{3} \and Oliver Hohlfeld\inst{1}}
\authorrunning{R\"uth et al.} 
\tocauthor{Jan R\"uth, Ingmar Poese, Christoph Dietzel, Oliver Hohlfeld}
\institute{RWTH Aachen University \email{\{rueth,hohlfeld\}@comsys.rwth-aachen.de}
\and Benocs GmbH \email{ipoese@benocs.com}
\and TU Berlin / DE-CIX \email{christoph@inet.tu-berlin.de}} 

\maketitle              

\copyrightnotice
\begin{abstract}
For the first time since the establishment of TCP and UDP, the Internet transport layer is subject to a major change by the introduction of QUIC.
Initiated by Google in 2012, QUIC provides a reliable, connection-oriented low-latency and fully encrypted transport.
In this paper, we provide the first broad assessment of QUIC usage in the wild.
We monitor the entire IPv4 address space since August 2016 and about 46\% of the DNS namespace to detected QUIC-capable infrastructures.
Our scans show that the number of QUIC-capable IPs has more than tripled since then to over \TotalIPsOctOneSeventeen{}.
We find around \Totaldomains{} domains hosted on QUIC-enabled infrastructure, but only 15K of them present valid certificates over QUIC.
Second, we analyze one year of traffic traces provided by MAWI, one day of a major European tier-1 ISP and from a large IXP to understand the dominance of QUIC in the Internet traffic mix.
We find QUIC to account for \DeCIXTotalQUICAVGSharePercent{} to \MobileTotalQUICAVGSharePercent{} of the current Internet traffic, depending on the vantage point.
This share is dominated by Google pushing up to \GoogleQUICPeakSharePercent{} of its traffic via QUIC.
\end{abstract}

\section{Introduction}
\label{sec:intro}
Recent years have fostered the understanding that TCP as the de-facto default Internet transport layer protocol has become a technological bottleneck that is hard to update.
This understanding is rooted in the fact that optimizing throughput is no longer a key concern in the Internet, but optimizing latency and providing encryption at the {\em transport} has become a concern.
The focus on latency results from shifted demands (e.g., by interactive web applications) and is currently proposed to be addressed in part by TCP extensions at the protocol level, e.g., TCP Fast Open~\cite{fastopen2011} or Multipath TCP~\cite{multipathTCP}.
While optimizing latency there is an additional demand to also provide an encrypted transport, typically realized by TLS on top of TCP.
Since this additional encryption adds additional latency, further optimizations address this latency inflation, e.g., 0-RTT in the upcoming TLS 1.3 standard~\cite{draft-tls13}.
While these approaches present clear advantages, their deployment is currently challenged by middleboxes and legacy systems.

Google's Quick UDP Internet Connections (QUIC) protocol~\cite{sigcommQUIC2017} aims to address these shortcomings in a new way.
Like TCP, it provides a connection-oriented, reliable, and in-order byte stream.
Yet unlike TCP, it enables stream multiplexing over a single connection while optimizing for latency.
By fully encrypting already at the transport layer, QUIC provides security and excludes (interfering) middlebox optimizations; thereby paving the way for a rapidly evolving transport layer.
By implementing QUIC in user space on top of UDP, its ability to rapidly update and customize a transport per application has yet unknown consequences and motives measurements.
It was first introduced to Chromium in 2012 and has undergone rapid development and high update-rate since then---as we will partly show in our measurements.
Since 2016, the IETF QUIC working group~\cite{QuicWG} is working on its standardization.
Google widely enabled QUIC for {\em all} of its users in January 2017~\cite{ietf96-quic-slides, sigcommQUIC2017}, motivating our study capturing its first 9 months of general deployment.
Yet, in contrast to TCP and TLS, there is very limited tool support to analyze QUIC and the academic understanding is currently limited to protocol security~\cite{fischlin2014, lychev2015, jager2015} and performance~\cite{carlucci2015, cook2017, sigcommQUIC2017, imcQUIC2017}. 

In this paper, we complement these works by providing the first large-scale analysis of the current QUIC {\em deployments} and its {\em traffic share}.
To assess the QUIC deployment, we regularly probe the entire IPv4 space for QUIC support since August 2016.
In our scans, we observe a growing adoption on QUIC reaching \TotalIPsOctOneSeventeen{} IPs supporting QUIC in October 2017, of which \GoogleIPsShareOctOneSeventeen{} (\AkamaiIPsShareOctOneSeventeen{}) are operated by Google (Akamai).
We additionally probe the complete set of .com/.net/.org domains as well as the Alexa Top 1M list, i.e., around 46\% of the domain name space~\cite{verisignReport}.
To assess the traffic share that these deployments generate, we analyzed traffic traces from three vantage points: {\em i)} 9 months of traffic in 2017 on a transit link to an ISP (MAWI dataset~\cite{mawi}), {\em ii)} one day in August 2017 at a European tier-1 ISP, representing edge (DSL + cellular) and backbone traffic, and {\em iii)} one day in August 2017 at a large European IXP.
In these networks, QUIC accounts for \DeCIXTotalQUICAVGSharePercent{} -- \MobileTotalQUICAVGSharePercent{} of the monitored traffic.
The observed traffic is largely contributed by Google (up to \GoogleQUICFromTotalQUICSharePercent{} in the ISP) and only marginally by Akamai (\AkamaiQUICFromTotalQUICSharePercent{} in the ISP and \DeCIXAkamaiQUICFromTotalQUICSharePercent{} in the IXP), despite having a large number of QUIC-capable IPs.
Our contributions are as follows.
\begin{itemize}\vspace{-.5em}
\item We analyze the development and deployment of QUIC in the IPv4 Internet.
\item We present the first view on QUIC deployment and traffic outside of Google's network from three different vantage points.
\item We build and together with this paper publish tools to: Enumerate QUIC hosts and tools to massively grab and decode QUIC protocol parameters.
\item We publish all our active measurement data and future scans on~\cite{DataAndSources}.
\end{itemize}
\vspace{-0.5em}
\afblock{Structure.} Section~\ref{sec:background} introduces the QUIC handshake as a basis for our host enumeration. 
Section~\ref{sec:scans} presents our view on QUIC in IPv4 and in three large TLDs as well as the tools that drive our measurements.
Section~\ref{sec:traffic} shows how QUIC reshapes traffic in local and ISP/IXP networks.
Section~\ref{sec:rw} discusses related works and Section~\ref{sec:discussion} concludes the paper.

\section{An Introduction to QUIC's Handshake}
\label{sec:background}
We first introduce the QUIC connection establishment phase that we utilize in our measurements for host enumeration and certificate grabbing.
For a broader discussion of QUIC's features and design choices we refer to~\cite{sigcommQUIC2017}.
We focus on the QUIC early deployment draft as the IETF draft is not yet fully specified.

\begin{figure}[t]
\center
\includegraphics{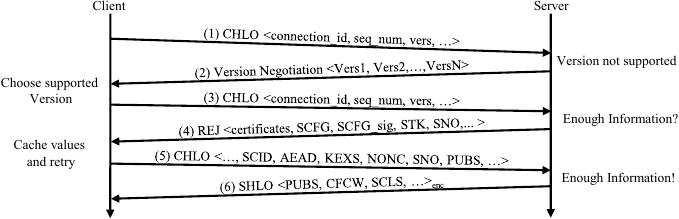}
\vspace{-1.3em}
\caption{A long QUIC handshake including version negotiation and caching of values.}
\label{fig:handshake}
\vspace{-1.5em}
\end{figure}

One of QUIC's main features is a fast connection establishment: In the ideal case, when cached information of a prior connection is available, it does not even take a single round-trip (0-RTT) to send encrypted application data.
Yet, in the worst case (without prior connections as in our measurements), QUIC needs at least three round-trips as shown in Figure~\ref{fig:handshake} and explained next.

Clients initiate a connection using a Client Hello (\texttt{CHLO})(1) including the QUIC version it desires to use.
In case the server does not support this version, it may send a version negotiation packet (2) enabling the client to choose from a list of supported versions for a second try.
We will utilize packet (1) to quickly probe for QUIC-capable hosts with only a single packet exchange and analyze their supported versions provided in (2).
Using a supported version, the client may advance in the handshake by sending another \texttt{CHLO} (3), without prior communication, it does not possess enough information about the server to establish a valid connection.
The server supplies the necessary information (4), in one or multiple exchanges (i.e., (3) and (4) may be repeated until all required data is available).
In these step(s), the client will be given a signed server config (\texttt{SCFG}) including supported ciphers, key exchange algorithms and their public values, and among other things the certificates authenticating the host.
We will utilize these information to analyze the server-provided certificates.
With this information, the client can issue another \texttt{CHLO} (5) including enough information to establish a connection, the client may even send encrypted data following the \texttt{CHLO} which depicts the optimal case for a 0-RTT connection establishment.
Following the \texttt{CHLO}, the server acknowledges (6) the successful connection establishment with a Server Hello (\texttt{SHLO}), containing further key/value-pairs enabling to fully utilize the connection.

\section{Availability: QUIC Server Infrastructures}
\label{sec:scans}
We start by analyzing the availability of QUIC in the Internet, i.e., how many IPs, domains, and infrastructures support QUIC.
If not stated otherwise, the results are based on scan data obtained in the first week of October 2017.

\subsection{Enumerating QUIC IPv4 Hosts}
\label{sec:hostscan}
\afblock{IP Scan Methodology.}
To quickly probe the entire IPv4 space for QUIC capable hosts, we extend ZMap~\cite{Durumeric13}, which enables to rapidly enumerate IPv4 addresses.
To identify QUIC hosts, we use QUIC's version negotiation feature (see Section~\ref{sec:background}).
As QUIC is build to enable rapid protocol development and deployment, negotiation of a supported version (i.e., supported by client and server) is fundamental to its design.
That is, the protocol requires to announce a version identifier in the initial packet sent from the client to the server.
In case the version announced by the client is not supported by the server, it sends a version negotiation packet.
This packet lists all supported versions by the server, enabling the client to find a common version that is used in a subsequent handshake.
We leverage this feature and sent a valid handshake message containing a version that is likely to be \emph{unsupported} by the other party, i.e., by including a version that is not reserved and does not follow the current pattern.
In response, the server will not be able to continue the handshake as both versions do not match, thus, it will send a version negotiation packet containing a list of its supported versions.
Using an invalid version has the advantage that we not only enumerate valid QUIC hosts but also gain further insights about the server, namely the list of its supported versions.
We declare an IP as QUIC-capable, if we either receive a valid version negotiation packet or a QUIC public reset packet (comparable to a TCP RST).
We build and publish~\cite{DataAndSources} ZMap modules implementing this behavior enabling rapid enumeration of QUIC hosts in the IPv4 space.

\begin{figure}[t]
\includegraphics{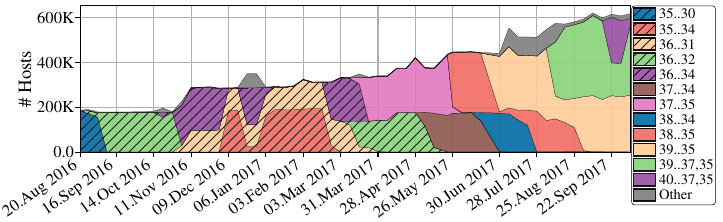}
\vspace{-2em}
\caption{Number of QUIC-capable IPs and support for sets of certain QUIC versions, here we display versions when there was support by at least 20000 hosts once. Versions that first appeared in 2016 are hatched.}
\label{fig:quic_overall}
\vspace{-1.5em}
\end{figure}

\afblock{QUIC Hosts.}
Figure~\ref{fig:quic_overall} shows that the total number of QUIC-capable IPs (sum of stacked area) has more than tripled from \TotalIPsAugOneSixteen{} IPs in August 2016 to \TotalIPsOctOneSeventeen{} IPs in October 2017.
As of October, we find IPs in \TotalASsOctOneSeventeen{} Autonomous Systems (ASs).
To analyze who drives this trend, we attribute QUIC IPs to providers:
we classify IPs by {\em i)} AS information, {\em ii)} per-IP X509 certificate data (e.g., who issued the certificate, who owns it), and {\em iii)} per-IP reverse DNS data (e.g., Akamai configures rDNS entries such as *.deploy.static.akamaitechnologies.com), using data available at Routeviews and scans.io.
As of August 2016, we can already attribute \GoogleIPsAugOneSixteen{} IPs to Google.
They have since doubled their QUIC-capable infrastructure to \GoogleIPsOctOneSeventeen{} IPs as of October 2017, accounting for \GoogleIPsShareOctOneSeventeen{} of all QUIC-capable IPs.
We identify Akamai as the second largest QUIC-enabler: they started to increasingly deploy QUIC on their servers in November 2016, while we find around \AkamaiIPsAugOneSixteen{} Akamai IPs in August, the number jumps to \AkamaiIPsNovOneSixteen{} IPs in November 2016.
Akamai has since then continued to deploy QUIC having \AkamaiIPsOctOneSeventeen{} IPs as of October 2017 accounting for \AkamaiIPsShareOctOneSeventeen{} of all QUIC-enabled IPs.

To classify the remaining \UnclassifiedIPsOctOneSeventeen{} hosts, we executed TCP HTTP GET / on port 80 for these IPs.
However, for \HTTPGETErrorOne{} IPs we could not get any data due to  \HTTPGETErrorNameOne s.
Apart from this, we find \HTTPGETServerOne{} hosts announcing a \textit{\HTTPGETServerNameOne{}} server string, a web server that added QUIC support in mid of July 2017~\cite{lightspeed}.
We find servers announcing \textit{\HTTPGETServerNameTwo{}} (\HTTPGETServerTwo{}) and \textit{\HTTPGETServerNameThree{}} (\HTTPGETServerThree{}), hinting at even more Google and Akamai installations.
The fourth largest group of servers announces \textit{\HTTPGETServerNameFour{}} (\HTTPGETServerFour{}) as the server string, this server uses the quic-go~\cite{quic-go} library and can also be used as a reverse proxy for other TCP-only servers.

\takeaway{We observe a steady growth of QUIC-capable IPs, mainly driven by Google and Akamai. Few IPs already use third-party server implementations.}

\afblock{QUIC Version Support.}
Since QUIC is under active development, it requires clients and servers to be regularly updated to support recent versions.
To understand how the server infrastructure is updated, Figure~\ref{fig:quic_overall} shows the number of hosts supporting a certain set of versions (recall: A host may support multiple versions!).
The figure shows that many version combinations have a short lifespan in which old versions fade away and new versions appear.
For example, hosts supporting version \texttt{Q035} down to version \texttt{Q030} switch to versions \texttt{Q036,...,Q032}, thus losing support for two versions.
Yet, while some versions fade away, we also see that, e.g., version \texttt{Q035} is supported by almost all hosts over the course of our dataset.
Even though, to the end of our observations support for version \texttt{Q036} is dropped.
While this shows that some versions offer a long-term support, the figure also shows how vibrant the QUIC landscape is.

Given that some versions introduce radical protocol changes without backward compatibility, questions concerning the long-term stability of a QUIC-Internet are raised.
On the one hand, the ability to easily update the protocol offers the possibility to quickly introduce new features and thereby to evolve the protocol.
On the other hand, updating Internet systems is known to be notoriously hard.
The vast amount of legacy systems raises the question of long-term compatibility---designing implementations to be easy to update is challenging. 

\takeaway{QUIC is currently subject to rapid development reflected in frequent version updates.
Given its realization in user space at the application-layer, this property is likely to stay: future transports can be potentially updated as frequently as any other application.
This motivates future measurements to assess the potentially highly dynamic future Internet transport landscape.}

\subsection{Enumerating QUIC Domain Names}
\afblock{Methodology.}
We develop a second tool that finishes the handshake and enables to further classify previously identified hosts and infrastructures.
To account for mandatory Server Name Indication (SNI), it can present a hostname that is necessary for the server to deliver correct certificates when hosting multiple sites on a single server.
We base our tool~\cite{DataAndSources} on the quic-go~\cite{quic-go} library which we extended to enable tracing
within the connection establishment to extract all handshake parameters (see Figure~\ref{fig:handshake}).

\begin{figure}[t]
\includegraphics{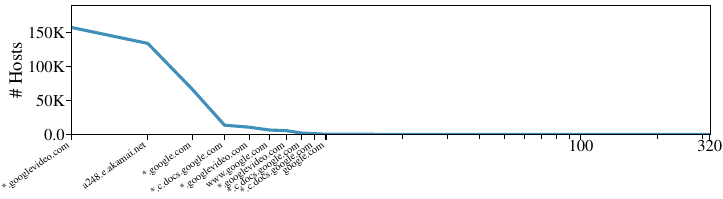}
\vspace{-2em}
\caption{Number of hosts giving out the same certificate on the y-axis. First listed common names for the 10 certificates with the highest coverage shown on the log x-axis.}
\label{fig:certs}
\vspace{-2em}
\end{figure}

\afblock{IP-based Certificate Scan.}
In a first step, we cluster all QUIC-enabled IPs discovered in Section~\ref{sec:hostscan} by their X509 certificate hash.
This step enables to better understand QUIC-enabled infrastructures.
Since the server's hostname is unknown at the request time when enumerating the IPv4 address space, we present dummy domains (e.g., foo.com) to each IP and retrieve the X509 certificate.
The retrieved certificate provides information on the domain names for which the certificate is valid, which can indicate the hosting infrastructure.
We remark that this approach yields the {\em default} website that is configured at a server and will not identify different sites in the presence of SNI.
In fact, we find that \CERTNumSNIRequired{} hosts require SNI and do not deliver a certificate (for which we account for when scanning domain zones later).
Figure~\ref{fig:certs} shows that we only observe \CERTNumUniqCerts{} different certificates for the probed \CERTNumProbedHosts{} QUIC IPs.
The heavy-tailed distribution shows the top-five (ten) certificates already represent \CERTShareTopFive{} (\CERTShareTopTen{}) of the IPs, most prominently Google and Akamai.
We validated that these IPs actually belong to both companies by requesting content via TCP and HTTP on port 80 on the same hosts.
We next assess QUIC support among domain names.

\afblock{Probing complete domain lists.}
Presenting a non-existing SNI name in our previous measurement will miss any server that enforces to present a valid hostname, thus we next assess the QUIC support by probing complete domain name lists.
That is, we probe all domains in the .com/.net/.org zone files and in the Alexa Top 1M list.
These zones are available at Verisign~\cite{verisignTLD} (.com/.net) and PIR~\cite{pir} (.org).
Together they contain more than 150M domains, i.e., about 46\,\% of the domain space~\cite{verisignReport}.
We use zDNS to resolve the domains and for each successful resolution, we use our tool to check for QUIC support and to grab all parameters from the connection establishment.
The whole process takes roughly 15\,h and is thus feasible to run on a daily basis.
Yet, as QUIC CHLO packets are padded to nearly fill the MTU, the scan easily saturates a 1Gbit link.

Table~\ref{tab:quic-tld} shows the QUIC-support in the .com/.net/.org zones as well as in the Alexa Top 1M list.
We define \textit{QUIC-enabled} domains as being able to initiate a QUIC handshake.
A domain is tagged as \textit{Timeout} when we received no response to our initial QUIC \texttt{CHLO} within 12 seconds, e.g., in the absence of QUIC support.
We furthermore show some specific errors as well as \textit{DNS-failures}.

\begin{table}[b]
	\vspace{-2em}
	\centering
	\resizebox{\linewidth}{!}{
\begin{tabular}{r ? r r ? r r ? r r ? r r }
        	\Xhline{0.75pt}
        	\multicolumn{1}{c}{}					& \multicolumn{2}{c}{\comDate}					& \multicolumn{2}{c}{\netDate}					& \multicolumn{2}{c}{\orgDate} 					& \multicolumn{2}{c}{\alexaDate} 	\\    

        \multicolumn{1}{c}{} 					& \multicolumn{2}{c}{\textbf{.com}} 	 			& \multicolumn{2}{c}{\textbf{.net}}				& \multicolumn{2}{c}{\textbf{.org}} 				& \multicolumn{2}{c}{\textbf{Alexa 1M}} 	\\    
	\Xhline{0.75pt}
        \textbf{\#\,Domains}					& \comTotal 		& (\comTotalPercent)			& \netTotal 		& (\netTotalPercent) 			& \orgTotal 		& (\orgTotalPercent) 			&\alexaTotal 			& (\alexaTotalPercent)		\\
        \hline
        \rowcolor{gray!25}\textbf{QUIC-enabled} 	& \comSupport 		&(\comSupportPercent)		& \netSupport 		& (\netSupportPercent)		& \orgSupport 		& (\orgSupportPercent) 		& \alexaSupport 		& (\alexaSupportPercent) 			\\
	\textbf{Valid Certificate} 					& \comTrueSupport 	&(\comTrueSupportPercent)	& \netTrueSupport 	& (\netTrueSupportPercent)	& \orgTrueSupport	& (\orgTrueSupportPercent)	& \alexaTrueSupport		& (\alexaTrueSupportPercent) 			\\
        \rowcolor{gray!25}\textbf{Timeout} 		& \comTimeout 		& (\comTimeoutPercent)		& \netTimeout 		& (\netTimeoutPercent)		& \orgTimeout 		& (\orgTimeoutPercent)		& \alexaTimeout 		& (\alexaTimeoutPercent)			\\
        \hline
        \textbf{Version-failed}				 	& \comInvalidVersion	& (\comInvalidVersionPercent)	& \netInvalidVersion 	& (\netInvalidVersionPercent)	& \orgInvalidVersion 	& (\orgInvalidVersionPercent)	& \alexaInvalidVersion 	& (\alexaInvalidVersionPercent)			\\
        \rowcolor{gray!25}\textbf{Protocol-error} 	& \comProtocolError 	& (\comProtocolErrorPercent)	& \netProtocolError 	& (\netProtocolErrorPercent) 	& \orgProtocolError 	& (\orgProtocolErrorPercent) 	& \alexaProtocolError 	& (\alexaProtocolErrorPercent) 			\\
        \textbf{Invalid-IP} 					& \comInvalidIP 	& (\comInvalidIPPercent) 		& \netInvalidIP 		& (\netInvalidIPPercent)		& \orgInvalidIP 		& (\orgInvalidIPPercent)		& \alexaInvalidIP 		& (\alexaInvalidIPPercent)			\\
        \rowcolor{gray!25}\textbf{DNS-failure} 	& \comDNSErrors 	& (\comDNSErrorsPercent)	& \netDNSErrors 	& (\netDNSErrorsPercent)		& \orgDNSErrors 	& (\orgDNSErrorsPercent)		& \alexaDNSErrors 		& (\alexaDNSErrorsPercent)			\\
        	\Xhline{0.75pt}
    \end{tabular}
    }
    \vspace{0.05em}
    \caption{QUIC support in different TLDs and in the Alexa Top 1M list. Weekly data is available at \url{https://quic.comsys.rwth-aachen.de}.}
    \label{tab:quic-tld}
    \vspace{-1em}
\end{table}

Overall QUIC-support is very low.
Depending on the zone, \netSupportPercent{}---\comSupportPercent{} domains are hosted on QUIC-enabled hosts.
Only \comShareTrueFromSupport{} -- \orgShareTrueFromSupport{} of these domains present a valid X509 certificate.
This questions how many domains actually deliver content via QUIC.

\afblock{Landing Page Content.}
Websites can utilize different server configuration and even different server implementations for different protocols.
The successful establishment of QUIC connections does thus not imply that meaningful content is being served.
To assess how many QUIC-capable domains deliver content similar to their HTTP~1.1/2 counterparts, we instruct Google's QUIC test client (part of the Chromium source) to download their landing page via QUIC.
We then compare their content to their HTTP~1.1/2 counterparts which should be similar if these QUIC-capable domains are properly set up.
We disabled certificate checks to probe all capable domains.
Out of the probed \Totaldomains{} domains, \TotalquicNullByteDomains{} (\TotalquicNullByteDomainsPercent{}) return no data and \TotalquicDataDomains{} (\TotalquicDataDomainsPercent{}) $>1$kB via QUIC.
In case of the latter, \TotalagreeingContent{} domains (\TotalagreeingContentAkamai{} served by Akamai) do deliver content similar to their HTTP 1.1/2 counterparts.
We define similarity by structural HTML similarity (e.g., in the number of tags, links, images, scripts, ...) and require $>3$ metrics to agree to define a web page to be similar.
Domains delivering similar content over QUIC are thus in principle ready to be served by a QUIC-capable browser.
To be discovered by a Chrome browser, they, however, need to present an alternative service ({\texttt{alt\_srv}) header via TCP-based HTTP(S) pointing to their QUIC counterpart.
\TotalhasHttpsAltSvcDomains{} domains present this header via HTTPS (\TotalhasHttpsAltSvcDomainsGoogle{} hosted by Google and \TotalhasHttpsAltSvcDomainsAkamai{} by Akamai) and only \TotalhasHttpAltSvcDomains{} via HTTP.
Thus a large share of the domains would not be contacted by a Chrome browser even though QUIC support is in principle available.
The header further specifies the QUIC versions supported by the server, of which at measurement time Chrome requires QUIC version 39.
Only \TotalchromeSupportedSomainsGoogle{} domains present this version in their \texttt{alt\_srv} header, all hosted by Google.
We remark that our content analysis only regards {\em landing} pages and does not account for additional assets (e.g., images or videos).
Particularly CDNs offer dedicated products for media delivery, whose QUIC support can differ.
Assessing their QUIC support in detail thus provides an interesting angle for future work.

\takeaway{The limited number of X509 certificates retrieved in our IP-based scan hints at the small number of different providers currently using or experimenting with QUIC.
Furthermore, only a small fraction of the monitored domains are hosted on QUIC-capable infrastructures--an even smaller fraction can actually deliver valid certificates for the requested domains.
Regardless, of the certificate, many QUIC-enabled domains do deliver their pages via QUIC.
Yet in our measurements, many would not be contacted by a Chrome browser, either because of a non-present \texttt{alt\_srv} header or insufficient version support.
There is thus a big potential to increase QUIC support. 
We next study how this QUIC-support is reflected in actual traffic shares.
}

\section{Usage: QUIC Traffic Share}
\label{sec:traffic}
We quantify the QUIC traffic share by analyzing three traces representing different vantage points:
{\em i)} 9 months of traffic in 2017 on a transit link to an upstream ISP (MAWI dataset~\cite{mawi}),
{\em ii)} one day in August 2017 at a European tier-1 ISP, representing edge (DSL + cellular) and backbone traffic, and
{\em iii)} the same day at a large European IXP.

\afblock{Traffic Classification.}
We use protocol and port information to classify HTTPS (TCP port 443), HTTP (TCP port 80), and QUIC (UDP port 443).
We chose this classification since it is applicable to all of our traces: MAWI (PCAP header traces) and ISP + IXP (Netflow traces without protocol headers).
We remark that this classification can {\em i)} miss protocol traffic on non-standard ports and can {\em ii)} wrongly attribute other traffic on the monitored ports.
However, it still enables to report an upper bound on the protocol usage on standard ports.
We show the per-trace traffic shares in Table~\ref{tab:traffic_shares} which we discuss next.

\input{traffictable2}

\begin{figure}[t]%
\includegraphics{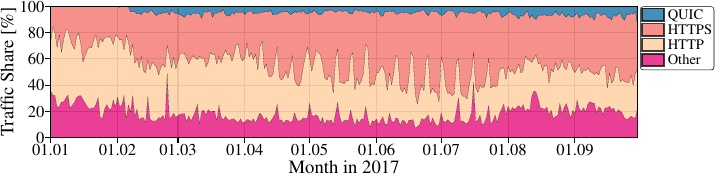}
\vspace{-2em}
\caption{Traffic share of QUIC compared to HTTP and HTTPS in the MAWI trace.}
\label{fig:mawi_share}
\vspace{-2em}
\end{figure}%

\afblock{MAWI Backbone Trace.}
We start by analyzing traffic on a trans-Pacific WIDE backbone link provided by the MAWI working group~\cite{mawi}.
We analyze anonymized header traces available at the MAWI repository (\textit{samplepoint~F}).
The monitored link is a transit link connecting the WIDE backbone to an upstream ISP.
The traces involve 15 minutes of traffic captured at 14h on each day.
Each packet is caped to the first 96~bytes.

We begin to analyze traffic on January 1\textsuperscript{st} 2017, since Google enabled QUIC for all of its Chrome and Google-developed Android App users in January 2017~\cite{sigcommQUIC2017}.
Figure~\ref{fig:mawi_share} shows the traffic volume until end of September 2017.
The trace shows that the QUIC traffic share is \MawiQUICAvgShareJan{} in January.
This is in contrast to the Google report of having widely enabled QUIC in January, suggesting that the monitored user-base is not using Google products (e.g., Chrome) at the time, QUIC has not been enabled for this network or that traffic is routed differently.
We observe the first QUIC traffic in February where the QUIC traffic share is at \MawiQUICAvgShareFeb{}.
It continues to increase to \MawiQUICAvgShareMar{} in March and reaches \MawiQUICAvgShareSep{} in September.
QUIC offers an alternative to TCP+TLS, which is the foundation of legacy HTTPS, its share is at around \MawiHTTPSAvgShareSep{}, even the unencrypted version HTTP is still at around \MawiHTTPAvgShareSep{}.
As the provided trace anonymizes destination and source addresses, we cannot attribute this traffic to infrastructures (e.g., Google or Akamai) or services (e.g., YouTube).
We leave this analysis to the ISP trace for which we have AS-level information available.

\takeaway{Within nine months after its general activation by Google, QUIC already accounts for a non-negligible traffic share, demonstrating its ability to evolve Internet transport.}

\afblock{European Tier-1 ISP.} 
We obtained anonymized and aggregated Netflow traces from all border routers of a large European ISP for one day in August 2017.
The Netflow traces were aggregated to 5-minute bins and all IP addresses were replaced by AS numbers before they were made available to us.
Thus the traces do not reveal the behavior of individual users.
The captured traffic contains {\em i)} edge traffic by DSL, {\em ii)} cellular customers, and {\em iii)} transit backbone traffic.

Figure~\ref{fig:isp_share} shows the traffic volume (up- and downstream) over the course of 24\,h by protocol and prominent infrastructures (the traffic volume (y-axis) has been removed at the request of the ISP).
As our previous host-based analysis (see Section~\ref{sec:scans}) showed that QUIC is mainly supported by Google and Akamai servers, we also show their traffic shares (according to their AS numbers).
At first, we observe that QUIC traffic follows the same daily pattern as HTTP and HTTPS.
On average QUIC accounts for \TotalQUICAVGSharePercent{} of the traffic with a standard deviation of $\sigma$:\,\TotalQUICSTDSharePercent.
This deviation is similar to HTTP ($\sigma$:\,\TotalHTTPSTDSharePercent) and HTTPS ($\sigma$:\,\TotalHTTPSSTDSharePercent) which account for \TotalHTTPAVGSharePercent{} and \TotalHTTPSAVGSharePercent{} of the traffic, respectively.

The observed QUIC traffic is almost exclusively contributed by Google: They account for \GoogleQUICFromTotalQUICSharePercent{} of the overall observed QUIC traffic.
Among all of Google's traffic, \GoogleQUICAVGSharePercent{} is using QUIC ($\sigma$:\,\GoogleQUICSTDSharePercent), peaking at \GoogleQUICPeakSharePercent{}.
This is a larger share than a global average of 32\,\% reported by Google in November 2016~\cite{sigcommQUIC2017}.
Currently, QUIC is mainly supported by Google-developed applications (e.g., Chrome or the Youtube Android app).
In the absence of QUIC libraries, third-party support is low (e.g., Opera has optional QUIC and Firefox no QUIC support).
The availability of QUIC libraries thus has the potential to drastically improve client support and therefore increase QUIC's traffic share.

\begin{figure}[t]
\includegraphics{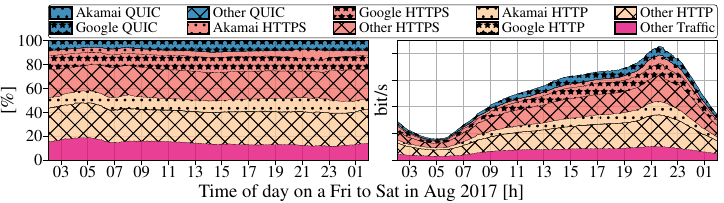}
\vspace{-2em}
\caption{QUIC traffic share in a major European ISP (up- and downstream). Left, relative share of QUIC. Right, total traffic compared to HTTP(S), y-axis has been anonymized at the request of the ISP. Nearly all QUIC traffic is served by Google. }
\label{fig:isp_share}
\vspace{-2.5em}
\end{figure}

\begin{figure}[t]
\includegraphics{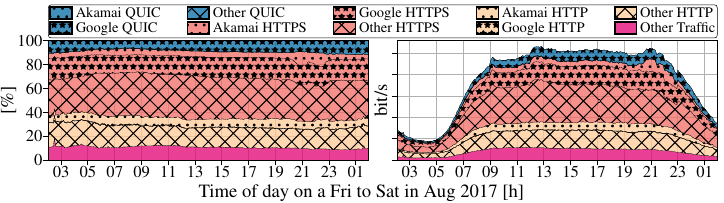}
\vspace{-2.25em}
\caption{Mobile network traffic share of QUIC in a major European ISP. Left, relative share of QUIC traffic. Right, absolute traffic share compared to HTTP(S), y-axis has been anonymized at the request of the ISP.}
\label{fig:isp_mobile_share}
\vspace{-1.25em}
\end{figure}

In contrast, Akamai only serves  \AkamaiQUICAVGSharePercent{}
 of its traffic via QUIC---despite contributing a large portion of the overall QUIC-capable IPs (\AkamaiIPsShareOctOneSeventeen{}, see Section~\ref{sec:hostscan}).
This discrepancy between the number of IPs and the traffic share suggests that QUIC is not yet widely activated among all customers/products.
Yet on average, Akamai accounts for \AkamaiHTTPFromTotalSharePercent{} (HTTP) and \AkamaiHTTPSFromTotalSharePercent{} (HTTPS) of all traffic and thus, together with the fact that they already have a QUIC-capable infrastructure, has the potential to shift more traffic towards QUIC.
A higher QUIC share has several implications, while QUIC and TCP are generally similar in nature, subtle differences in the protocols may influence the performance of whole networks, e.g., by default QUIC uses larger initial congestion windows than those standardized for TCP by IETF and demands pacing for smoothing the traffic.

\afblock{Mobile ISP.}
The ISP supplied us with information which traffic is for their mobile (cellular) customers, which we show in Figure~\ref{fig:isp_mobile_share}.
Please note that the reported mobile traffic is also contained in Figure~\ref{fig:isp_share}.
In contrast to the entire network of the ISP, the mobile traffic shows a different traffic pattern: while its throughput also decreases over night, mobile traffic rapidly increases in the morning and stays rather constant over the course of the day.
Apart from this, the average QUIC share in the mobile network of \MobileTotalQUICAVGSharePercent{} ($\sigma$: \MobileTotalQUICSTDSharePercent{}), the highest share among all traces ( see Table~\ref{tab:traffic_shares}).
In contrast, among the entire mobile Google traffic, only \MobileGoogleQUICAVGSharePercent{} ($\sigma$: \MobileGoogleQUICSTDSharePercent{}) is served via QUIC, lower than overall for the ISP.
Also for mobile traffic, Akamai only serves a negligible share of its traffic via QUIC and thus has the potential increase the QUIC traffic share.

\takeaway{QUIC traffic shares do (yet) not reflect server support. While Akamai operates a comparably large infrastructure in the number of QUIC-capable IPs, QUIC traffic is (still) almost entirely served by Google: this is likely to change.}

\begin{figure}[t]
\includegraphics{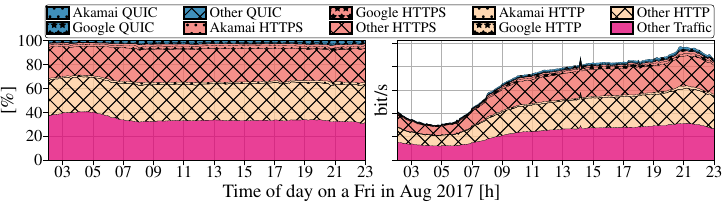}
\vspace{-2.25em}
\caption{QUIC traffic share at a large European IXP. Left, relative share of QUIC traffic. Right, absolute traffic share compared to HTTP(S), y-axis has been anonymized at the request of the IXP.}
\label{fig:ixp_traffic_share}
\vspace{-1.75em}
\end{figure}

\afblock{European IXP.}
We obtained sampled flow data of a large (European) IXP for the same day in August as for the ISP and show its traffic share in Figure~\ref{fig:ixp_traffic_share}.
We classify Google and Akamai traffic by customer port information---since both peer at the IXP---and plot their HTTP(S) and QUIC traffic shares similar to the ISP.
On average QUIC accounts for \DeCIXTotalQUICAVGSharePercent{} ($\sigma$:\,\TotalQUICSTDSharePercent{}) of the traffic, which is lowest share among all traces (see Tables~\ref{tab:traffic_shares}).
Unlike the ISP, the largest portion is now contributed by Akamai (\DeCIXAkamaiQUICFromTotalQUICSharePercent{}) and we observe a lower share of Google traffic (\DeCIXGoogleQUICFromTotalQUICSharePercent{})---recall that Google contributed \GoogleQUICFromTotalQUICSharePercent{} of the QUIC traffic at the ISP.

\takeaway{(Per-CDN) traffic shares largely depend on the chosen vantage point.} 

\afblock{Discussion.}%
We observe different QUIC traffic shares at the ISP/IXP and particularly different shares of the QUIC traffic by Google / Akamai (relative to the overall traffic of each vantage point). 
These vantage point dependent differences are likely caused by different traffic engineering strategies since both providers peer at both vantage points. 
These differences highlight that observed traffic shares are in general highly vantage point dependent. 
Understanding the incentives for these different traffic engineering strategies is an interesting starting point for future research.

\section{Related Work}
\label{sec:rw}

\afblock{QUIC Security.}
A first security analysis QUIC's key exchange is presented in~\cite{fischlin2014}, followed by a later analysis of the complete protocol~\cite{lychev2015}.
These works on the security analysis are complemented by presenting an attack vector in which the server config can be computed offline to impersonate the server~\cite{jager2015}.

\afblock{QUIC Performance.}
An early performance comparison of QUIC and HTTP1~\cite{carlucci2015} indicates that QUIC can provide better goodput and lower page loading times as traditional HTTP1 over TCP.
A more extensive evaluation in~\cite{cook2017} also involves the comparison to HTTP2 and shows that QUIC can outperform HTTP2 over TCP/TLS, a finding that is supported by extensive evaluations in~\cite{imcQUIC2017}.
The reported performance experience by Google~\cite{sigcommQUIC2017} shows that QUIC lowers the Google search latency by 3.6--8\% and reduces YouTube rebuffering by 15--18\%.

We complement these works by providing the first broad assessment of QUIC usage in the wild and outside Google's network.
We study both the QUIC-enabled infrastructures and its traffic shares from three vantage points.

\section{Discussion and Conclusion}
\label{sec:discussion}

This paper presents the first broad assessment of QUIC, nine months after the general activation by Google for all of its users.
We study both the available {\em infrastructure} in terms of the number of QUIC-capable IPs and domains and their {\em traffic share} at three vantage points.
By probing the entire IPv4 address space, we find a steadily growing number of QUIC-enabled IPs which has tripled since August 2016 and reached  \TotalIPsOctOneSeventeen{} in October 2017.
This growth is mainly driven by Google and Akamai, which account for \GoogleIPsShareOctOneSeventeen{} and \AkamaiIPsShareOctOneSeventeen{} of these IPs. 
When regularly probing $\approx 150$M domains for QUIC support, we observe \Totaldomains{} capable domains of which \TotalagreeingContent{} serve content similar to their HTTP1/2 counterparts and only 15K present valid certificates.
Many (of the non-Google hosted) domains would not be contacted by a Chrome browser, either because of a non-present alternative service headers in HTTP(S) or insufficient version support.
This infrastructure size does, however, not reflect their traffic share: depending on the vantage point, Google accounts for \GoogleQUICFromTotalQUICSharePercent{} (ISP) of the QUIC traffic and Akamai contributes \AkamaiQUICFromTotalQUICSharePercent{} (ISP) to \DeCIXAkamaiQUICFromTotalQUICSharePercent{} (IXP), despite operating a similarly large number of QUIC-capable IPs. 
Given the factors that impede QUIC support, the QUIC traffic share is likely to increase in the future when being largely enabled at a wide range of infrastructures.

Realized as user space application-layer protocol, QUIC paves the way towards a rapidly evolving transport that can be updated as easily and as frequently as any application.
This aspect is manifested in the short lifetime of QUIC versions observed in our measurements while the protocol is still under development.
In light of these findings we expect a highly dynamic future Internet transport landscape to be studied and observed by future work.
\vspace{-0.5em}
\section*{Acknowledgments}
\vspace{-1em}
This work has been funded by the DFG as part of the CRC 1053 MAKI and SPP 1914 REFLEXES, and by European Union’s Horizon 2020 research and innovation programme under the ENDEAVOUR project (grant agreement 644960).
We would like to thank the network operators at RWTH Aachen University, especially Jens Hektor and Bernd Kohler.
We further thank our shepherd Tobias Flach and the anonymous reviewers.
Furthermore, we would like to thank Konrad Wolsing for maintaining our changes to the quic-go implementation.

\vspace{-0.5em}
\bibliographystyle{abbrv_1etal}
\bibliography{literature.bib}

\end{document}

%% file: quic-tlds.tex
\def\comTotal{{129.36\,M}}

\def\comTotalPercent{{100.0\%}}

\def\comSupport{{133.63\,K}}

\def\comSupportPercent{{0.1\%}}

\def\comInvalidIP{{322.24\,K}}

\def\comInvalidIPPercent{{0.25\%}}

\def\comTimeout{{114.63\,M}}

\def\comTimeoutPercent{{88.61\%}}

\def\comProtocolError{{606}}

\def\comProtocolErrorPercent{{0.0\%}}

\def\comDNSErrors{{13.76\,M}}

\def\comDNSErrorsPercent{{10.64\%}}

\def\comInvalidVersion{{29}}

\def\comInvalidVersionPercent{{0.0\%}}

\def\comTrueSupport{{2.14\,K}}

\def\comTrueSupportPercent{{0.0\%}}

\def\comShareTrueFromSupport{{1.6\%}}

\def\comDate{{06. Oct 2017}}

\def\orgTotal{{10.37\,M}}

\def\orgTotalPercent{{100.0\%}}

\def\orgSupport{{6.51\,K}}

\def\orgSupportPercent{{0.06\%}}

\def\orgInvalidIP{{40.15\,K}}

\def\orgInvalidIPPercent{{0.39\%}}

\def\orgTimeout{{8.09\,M}}

\def\orgTimeoutPercent{{78.06\%}}

\def\orgProtocolError{{0}}

\def\orgProtocolErrorPercent{{0.0\%}}

\def\orgDNSErrors{{1.18\,M}}

\def\orgDNSErrorsPercent{{11.41\%}}

\def\orgInvalidVersion{{1}}

\def\orgInvalidVersionPercent{{0.0\%}}

\def\orgTrueSupport{{159}}

\def\orgTrueSupportPercent{{0.0\%}}

\def\orgShareTrueFromSupport{{2.44\%}}

\def\orgDate{{04. Oct 2017}}

\def\netTotal{{14.75\,M}}

\def\netTotalPercent{{100.0\%}}

\def\netSupport{{8.73\,K}}

\def\netSupportPercent{{0.06\%}}

\def\netInvalidIP{{59.24\,K}}

\def\netInvalidIPPercent{{0.4\%}}

\def\netTimeout{{10.80\,M}}

\def\netTimeoutPercent{{73.23\%}}

\def\netProtocolError{{222}}

\def\netProtocolErrorPercent{{0.0\%}}

\def\netDNSErrors{{2.40\,M}}

\def\netDNSErrorsPercent{{16.26\%}}

\def\netInvalidVersion{{6}}

\def\netInvalidVersionPercent{{0.0\%}}

\def\netTrueSupport{{181}}

\def\netTrueSupportPercent{{0.0\%}}

\def\netDate{{03. Oct 2017}}

\def\alexaTotal{{999.94\,K}}

\def\alexaTotalPercent{{100.0\%}}

\def\alexaSupport{{11.97\,K}}

\def\alexaSupportPercent{{1.2\%}}

\def\alexaInvalidIP{{15.42\,K}}

\def\alexaInvalidIPPercent{{1.54\%}}

\def\alexaTimeout{{826.67\,K}}

\def\alexaTimeoutPercent{{82.67\%}}

\def\alexaProtocolError{{1}}

\def\alexaProtocolErrorPercent{{0.0\%}}

\def\alexaDNSErrors{{49.34\,K}}

\def\alexaDNSErrorsPercent{{4.93\%}}

\def\alexaInvalidVersion{{5}}

\def\alexaInvalidVersionPercent{{0.0\%}}

\def\alexaTrueSupport{{342}}

\def\alexaTrueSupportPercent{{0.03\%}}

\def\alexaDate{{08. Oct 2017}}

%% file: quic-isp-shares.tex
\def\GoogleHTTPAVGSharePercent{{1.4\%}}

\def\GoogleHTTPSAVGSharePercent{{59.5\%}}

\def\GoogleQUICAVGSharePercent{{39.1\%}}

\def\GoogleQUICPeakSharePercent{{42.1\%}}

\def\GoogleQUICSTDSharePercent{{2.3\%}}

\def\AkamaiHTTPAVGSharePercent{{67.9\%}}

\def\AkamaiHTTPSAVGSharePercent{{32.1\%}}

\def\AkamaiQUICAVGSharePercent{{0.1\%}}

\def\TotalHTTPAVGSharePercent{{37.7\%}}

\def\TotalHTTPSAVGSharePercent{{40.1\%}}

\def\TotalQUICAVGSharePercent{{7.8\%}}

\def\TotalHTTPSTDSharePercent{{1.2\%}}

\def\TotalHTTPSSTDSharePercent{{1.4\%}}

\def\TotalQUICSTDSharePercent{{1.0\%}}

\def\AkamaiQUICFromTotalQUICSharePercent{{0.1\%}}

\def\GoogleQUICFromTotalQUICSharePercent{{98.1\%}}

\def\AkamaiHTTPFromTotalHTTPSharePercent{{27.2\%}}

\def\GoogleHTTPFromTotalHTTPSharePercent{{0.7\%}}

\def\AkamaiHTTPSFromTotalHTTPSSharePercent{{12.6\%}}

\def\GoogleHTTPSFromTotalHTTPSSharePercent{{28.8\%}}

\def\AkamaiHTTPFromTotalSharePercent{{10.3\%}}

\def\AkamaiHTTPSFromTotalSharePercent{{5.1\%}}

%% file: quic-Mobileisp-shares.tex
\def\MobileGoogleHTTPAVGSharePercent{{1.6\%}}

\def\MobileGoogleHTTPSAVGSharePercent{{64.4\%}}

\def\MobileGoogleQUICAVGSharePercent{{34.0\%}}

\def\MobileGoogleQUICSTDSharePercent{{2.6\%}}

\def\MobileAkamaiHTTPAVGSharePercent{{57.7\%}}

\def\MobileAkamaiHTTPSAVGSharePercent{{42.3\%}}

\def\MobileAkamaiQUICAVGSharePercent{{0.0\%}}

\def\MobileTotalHTTPAVGSharePercent{{24.8\%}}

\def\MobileTotalHTTPSAVGSharePercent{{55.4\%}}

\def\MobileTotalQUICAVGSharePercent{{9.1\%}}

\def\MobileTotalQUICSTDSharePercent{{1.4\%}}

\def\MobileAkamaiQUICFromTotalQUICSharePercent{{0.1\%}}

\def\MobileGoogleQUICFromTotalQUICSharePercent{{96.9\%}}

\def\MobileAkamaiHTTPFromTotalHTTPSharePercent{{28.5\%}}

\def\MobileGoogleHTTPFromTotalHTTPSharePercent{{1.8\%}}

\def\MobileAkamaiHTTPSFromTotalHTTPSSharePercent{{9.6\%}}

\def\MobileGoogleHTTPSFromTotalHTTPSSharePercent{{29.5\%}}

%% file: quic-mawi-shares.tex
\def\MawiQUICAvgShareMar{{5.2\%}}

\def\MawiQUICAvgShareFeb{{3.9\%}}

\def\MawiHTTPAvgShareSep{{28.0\%}}

\def\MawiHTTPSAvgShareSep{{44.9\%}}

\def\MawiQUICAvgShareSep{{6.7\%}}

\def\MawiQUICAvgShareJan{{0.0\%}}

%% file: quic-asinfos.tex
\def\TotalIPsAugOneSixteen{{186.77\,K}}

\def\AkamaiIPsAugOneSixteen{{983}}

\def\GoogleIPsAugOneSixteen{{169.52\,K}}

\def\AkamaiIPsNovOneSixteen{{44.47\,K}}

\def\TotalASsOctOneSeventeen{{3.04\,K}}

\def\TotalIPsOctOneSeventeen{{617.59\,K}}

\def\AkamaiIPsOctOneSeventeen{{251.43\,K}}

\def\GoogleIPsOctOneSeventeen{{330.62\,K}}

\def\UnclassifiedIPsOctOneSeventeen{{35.54\,K}}

\def\AkamaiIPsShareOctOneSeventeen{{40.71\%}}

\def\GoogleIPsShareOctOneSeventeen{{53.53\%}}

%% file: quic-latesthttpgrep.tex
\def\HTTPGETErrorOne{{23.91\,K}}

\def\HTTPGETErrorNameOne{{i/o timeout}}

\def\HTTPGETServerOne{{7.34\,K}}

\def\HTTPGETServerNameOne{{LiteSpeed}}

\def\HTTPGETServerTwo{{1.69\,K}}

\def\HTTPGETServerNameTwo{{gws}}

\def\HTTPGETServerThree{{1.44\,K}}

\def\HTTPGETServerNameThree{{AkamaiGHost}}

\def\HTTPGETServerFour{{356}}

\def\HTTPGETServerNameFour{{Caddy}}

%% file: quic-certificate-data.tex
\def\CERTNumSNIRequired{{216.64\,K}}

\def\CERTNumUniqCerts{{320}}

\def\CERTNumProbedHosts{{617.59\,K}}

\def\CERTShareTopFive{{95.41\%}}

\def\CERTShareTopTen{{99.28\%}}

%% file: quic-DeCIX-shares.tex
\def\DeCIXGoogleHTTPAVGSharePercent{{3.1\%}}

\def\DeCIXGoogleHTTPSAVGSharePercent{{70.0\%}}

\def\DeCIXGoogleQUICAVGSharePercent{{26.9\%}}

\def\DeCIXAkamaiHTTPAVGSharePercent{{33.3\%}}

\def\DeCIXAkamaiHTTPSAVGSharePercent{{33.3\%}}

\def\DeCIXAkamaiQUICAVGSharePercent{{33.3\%}}

\def\DeCIXTotalHTTPAVGSharePercent{{32.2\%}}

\def\DeCIXTotalHTTPSAVGSharePercent{{30.9\%}}

\def\DeCIXTotalQUICAVGSharePercent{{2.6\%}}

\def\DeCIXAkamaiQUICFromTotalQUICSharePercent{{59.9\%}}

\def\DeCIXGoogleQUICFromTotalQUICSharePercent{{33.1\%}}

\def\DeCIXAkamaiHTTPFromTotalHTTPSharePercent{{5.0\%}}

\def\DeCIXGoogleHTTPFromTotalHTTPSharePercent{{0.3\%}}

\def\DeCIXAkamaiHTTPSFromTotalHTTPSSharePercent{{5.2\%}}

\def\DeCIXGoogleHTTPSFromTotalHTTPSSharePercent{{7.2\%}}

%% file: content-valuesredef.tex
\def\comsupportchrome39SupportedDomainsAkamai{{20}}

\def\comsupportchrome39SupportedDomains{{10K}}

\def\comsupportchrome39SupportedSomainsGoogle{{4K}}

\def\comsupportchrome39SupportedDomainsAkamaiPercent{{0.0\%}}

\def\comsupportchrome39SupportedDomainsPercent{{7.2\%}}

\def\comsupportchrome39SupportedSomainsGooglePercent{{2.9\%}}

\def\comtruesupportchrome39SupportedDomainsAkamai{{0}}

\def\comtruesupportchrome39SupportedDomains{{69}}

\def\comtruesupportchrome39SupportedSomainsGoogle{{59}}

\def\comtruesupportchrome39SupportedDomainsAkamaiPercent{{0.0\%}}

\def\comtruesupportchrome39SupportedDomainsPercent{{3.2\%}}

\def\comtruesupportchrome39SupportedSomainsGooglePercent{{2.8\%}}

\def\netsupportchrome39SupportedDomainsAkamai{{0}}

\def\netsupportchrome39SupportedDomains{{772}}

\def\netsupportchrome39SupportedSomainsGoogle{{281}}

\def\netsupportchrome39SupportedDomainsAkamaiPercent{{0.0\%}}

\def\netsupportchrome39SupportedDomainsPercent{{8.8\%}}

\def\netsupportchrome39SupportedSomainsGooglePercent{{3.2\%}}

\def\nettruesupportchrome39SupportedDomainsAkamai{{0}}

\def\nettruesupportchrome39SupportedDomains{{4}}

\def\nettruesupportchrome39SupportedSomainsGoogle{{2}}

\def\nettruesupportchrome39SupportedDomainsAkamaiPercent{{0.0\%}}

\def\nettruesupportchrome39SupportedDomainsPercent{{2.2\%}}

\def\nettruesupportchrome39SupportedSomainsGooglePercent{{1.1\%}}

\def\alexasupportchrome39SupportedDomainsAkamai{{0}}

\def\alexasupportchrome39SupportedDomains{{471}}

\def\alexasupportchrome39SupportedSomainsGoogle{{206}}

\def\alexasupportchrome39SupportedDomainsAkamaiPercent{{0.0\%}}

\def\alexasupportchrome39SupportedDomainsPercent{{3.9\%}}

\def\alexasupportchrome39SupportedSomainsGooglePercent{{1.7\%}}

\def\alexatruesupportchrome39SupportedDomainsAkamai{{0}}

\def\alexatruesupportchrome39SupportedDomains{{219}}

\def\alexatruesupportchrome39SupportedSomainsGoogle{{22}}

\def\alexatruesupportchrome39SupportedDomainsAkamaiPercent{{0.0\%}}

\def\alexatruesupportchrome39SupportedDomainsPercent{{64.4\%}}

\def\alexatruesupportchrome39SupportedSomainsGooglePercent{{6.5\%}}

\def\orgsupportchrome39SupportedDomainsAkamai{{1}}

\def\orgsupportchrome39SupportedDomains{{491}}

\def\orgsupportchrome39SupportedSomainsGoogle{{239}}

\def\orgsupportchrome39SupportedDomainsAkamaiPercent{{0.0\%}}

\def\orgsupportchrome39SupportedDomainsPercent{{7.5\%}}

\def\orgsupportchrome39SupportedSomainsGooglePercent{{3.7\%}}

\def\orgtruesupportchrome39SupportedDomainsAkamai{{0}}

\def\orgtruesupportchrome39SupportedDomains{{10}}

\def\orgtruesupportchrome39SupportedSomainsGoogle{{7}}

\def\orgtruesupportchrome39SupportedDomainsAkamaiPercent{{0.0\%}}

\def\orgtruesupportchrome39SupportedDomainsPercent{{6.3\%}}

\def\orgtruesupportchrome39SupportedSomainsGooglePercent{{4.4\%}}

\def\TotalhasHttpAltSvcDomains{{7}}

\def\TotalhasHttpsAltSvcDomains{{11K}}

\def\Totalchrome39SupportedDomainsAkamai{{21}}

\def\TotalquicDataDomains{{33K}}

\def\TotalagreeingContent{{33K}}

\def\TotalagreeingContentAkamai{{22K}}

\def\Totalchrome39SupportedDomains{{11K}}

\def\TotalquicNullByteDomains{{16K}}

\def\TotalhasHttpsAltSvcDomainsGoogle{{5K}}

\def\Totaldomains{{161K}}

\def\TotalhasHttpsAltSvcDomainsAkamai{{0}}

\def\Totalchrome39SupportedSomainsGoogle{{5K}}

\def\Totalchrome39SupportedDomainsAkamaiPercent{{0.0\%}}

\def\TotalquicDataDomainsPercent{{20.7\%}}

\def\Totalchrome39SupportedDomainsPercent{{7.0\%}}

\def\TotalquicNullByteDomainsPercent{{9.8\%}}

\def\Totalchrome39SupportedSomainsGooglePercent{{2.9\%}}

\def\TotalchromeSupportedSomainsGoogle{{5K}}

%% file: traffictable2.tex

\begin{table}[t]
\vspace{-0.1em}
\setlength{\tabcolsep}{0.45em} 
\centering
\begin{tabular}{@{}crrrlrrrrrr@{}}
\toprule
\multicolumn{1}{l}{} & \multicolumn{3}{c}{\textbf{Overall}} &  & \multicolumn{3}{c}{\textbf{Operator's share}} & \multicolumn{3}{c}{\textbf{Share in Protocol}} \\
\multicolumn{1}{l|}{} & \multicolumn{1}{c}{ \scriptsize \textbf{ HTTP}} & \multicolumn{1}{c}{\scriptsize\textbf{HTTPS}} & \multicolumn{1}{c|}{\scriptsize\textbf{QUIC}} & \multicolumn{1}{l|}{} & \multicolumn{1}{c}{\scriptsize\textbf{HTTP}} & \multicolumn{1}{c}{\scriptsize\textbf{HTTPS}} & \multicolumn{1}{c|}{\scriptsize\textbf{QUIC}} & \multicolumn{1}{c}{\scriptsize\textbf{HTTP}} & \multicolumn{1}{c}{\scriptsize\textbf{HTTPS}} & \multicolumn{1}{c}{\scriptsize\textbf{QUIC}} \\ \midrule
\multicolumn{1}{c|}{\textbf{MAWI}} & \MawiHTTPAvgShareSep & \MawiHTTPSAvgShareSep & \multicolumn{1}{r|}{\MawiQUICAvgShareSep} & \multicolumn{1}{c|}{-} & \multicolumn{3}{c|}{-} & \multicolumn{3}{c}{-} \\ \midrule
\multicolumn{1}{c|}{\multirow{2}{*}{\textbf{ISP}}} & \multirow{2}{*}{\TotalHTTPAVGSharePercent} & \multirow{2}{*}{\TotalHTTPSAVGSharePercent} & \multicolumn{1}{r|}{\multirow{2}{*}{\TotalQUICAVGSharePercent}} & \multicolumn{1}{l|}{Akamai} & \AkamaiHTTPAVGSharePercent & \AkamaiHTTPSAVGSharePercent & \multicolumn{1}{r|}{\AkamaiQUICAVGSharePercent} & \AkamaiHTTPFromTotalHTTPSharePercent & \AkamaiHTTPSFromTotalHTTPSSharePercent &  \AkamaiQUICFromTotalQUICSharePercent \\
\multicolumn{1}{c|}{} &  &  & \multicolumn{1}{r|}{} & \multicolumn{1}{l|}{Google} & \GoogleHTTPAVGSharePercent & \GoogleHTTPSAVGSharePercent & \multicolumn{1}{r|}{\GoogleQUICAVGSharePercent} & \GoogleHTTPFromTotalHTTPSharePercent & \GoogleHTTPSFromTotalHTTPSSharePercent & \GoogleQUICFromTotalQUICSharePercent \\ \midrule
\multicolumn{1}{c|}{\multirow{2}{*}{\textbf{\begin{tabular}[c]{@{}c@{}}Mobile\\ ISP\end{tabular}}}} & \multirow{2}{*}{\MobileTotalHTTPAVGSharePercent} & \multirow{2}{*}{\MobileTotalHTTPSAVGSharePercent} & \multicolumn{1}{r|}{\multirow{2}{*}{\MobileTotalQUICAVGSharePercent}} & \multicolumn{1}{l|}{Akamai} & \MobileAkamaiHTTPAVGSharePercent & \MobileAkamaiHTTPSAVGSharePercent & \multicolumn{1}{r|}{\MobileAkamaiQUICAVGSharePercent} & \MobileAkamaiHTTPFromTotalHTTPSharePercent & \MobileAkamaiHTTPSFromTotalHTTPSSharePercent & \MobileAkamaiQUICFromTotalQUICSharePercent \\
\multicolumn{1}{c|}{} &  &  & \multicolumn{1}{r|}{} & \multicolumn{1}{l|}{Google} & \MobileGoogleHTTPAVGSharePercent & \MobileGoogleHTTPSAVGSharePercent & \multicolumn{1}{r|}{\MobileGoogleQUICAVGSharePercent} & \MobileGoogleHTTPFromTotalHTTPSharePercent & \MobileGoogleHTTPSFromTotalHTTPSSharePercent & \MobileGoogleQUICFromTotalQUICSharePercent \\ \midrule
\multicolumn{1}{c|}{\multirow{2}{*}{\textbf{IXP}}} & \multirow{2}{*}{\DeCIXTotalHTTPAVGSharePercent} & \multirow{2}{*}{\DeCIXTotalHTTPSAVGSharePercent} & \multicolumn{1}{r|}{\multirow{2}{*}{\DeCIXTotalQUICAVGSharePercent}} & \multicolumn{1}{l|}{Akamai} & \DeCIXAkamaiHTTPAVGSharePercent & \DeCIXAkamaiHTTPSAVGSharePercent & \multicolumn{1}{r|}{\DeCIXAkamaiQUICAVGSharePercent} & \DeCIXAkamaiHTTPFromTotalHTTPSharePercent & \DeCIXAkamaiHTTPSFromTotalHTTPSSharePercent & \DeCIXAkamaiQUICFromTotalQUICSharePercent \\
\multicolumn{1}{c|}{} &  &  & \multicolumn{1}{r|}{} & \multicolumn{1}{l|}{Google} & \DeCIXGoogleHTTPAVGSharePercent & \DeCIXGoogleHTTPSAVGSharePercent & \multicolumn{1}{r|}{\DeCIXGoogleQUICAVGSharePercent} & \DeCIXGoogleHTTPFromTotalHTTPSharePercent & \DeCIXGoogleHTTPSFromTotalHTTPSSharePercent & \DeCIXGoogleQUICFromTotalQUICSharePercent \\ \bottomrule
\end{tabular}
\vspace{0.05em}
\caption{Average traffic shares (overall), among the operators,  and among the protocol. Operator's share is e.g., from all of Google's traffic the share of the QUIC traffic at a vantage point. Share in Protocols denotes the traffic share of a protocol at a vantage point, e.g., the amount of Google QUIC traffic from all other QUIC traffic.}
\label{tab:traffic_shares}
\vspace{-2.5em}
\end{table}

%% file: paper.bbl
\begin{thebibliography}{10}

\bibitem{DataAndSources}
Active measurements and tools.
\newblock \url{https://quic.comsys.rwth-aachen.de}.

\bibitem{QuicWG}
{IETF QUIC} working group.
\newblock \url{https://datatracker.ietf.org/wg/quic/about/}.

\bibitem{carlucci2015}
G.~Carlucci, et~al.
\newblock {HTTP over UDP: An Experimental Investigation of QUIC}.
\newblock In {\em Proceedings of the 30th Annual ACM Symposium on Applied
  Computing}, 2015.

\bibitem{quic-go}
L.~Clemente.
\newblock quic-go.
\newblock \url{https://github.com/lucas-clemente/quic-go}.

\bibitem{cook2017}
S.~Cook, et~al.
\newblock {QUIC: Better For What and For Whom?}
\newblock In {\em IEEE ICC}, 2017.

\bibitem{Durumeric13}
Z.~Durumeric, et~al.
\newblock Zmap: Fast internet-wide scanning and its security applications.
\newblock In {\em USENIX Security}, 2013.

\bibitem{fischlin2014}
M.~Fischlin and F.~G\"{u}nther.
\newblock {Multi-Stage Key Exchange and the Case of Google's QUIC Protocol}.
\newblock In {\em ACM CCS}, 2014.

\bibitem{jager2015}
T.~Jager, et~al.
\newblock {On the Security of TLS 1.3 and QUIC Against Weaknesses in PKCS\#1
  V1.5 Encryption}.
\newblock In {\em ACM CCS}, 2015.

\bibitem{imcQUIC2017}
A.~M. Kakhki, et~al.
\newblock {Taking a Long Look at QUIC: An Approach for Rigorous Evaluation of
  Rapidly Evolving Transport Protocols}.
\newblock In {\em ACM IMC}, 2017.

\bibitem{sigcommQUIC2017}
A.~Langley, et~al.
\newblock {The QUIC Transport Protocol: Design and Internet-Scale Deployment}.
\newblock In {\em ACM SIGCOMM}, 2017.

\bibitem{lightspeed}
{LiteSpeed Technologies Inc.}
\newblock {LiteSpeed --- Release Log}.
\newblock
  \url{https://www.litespeedtech.com/products/litespeed-web-server/release-log}.

\bibitem{lychev2015}
R.~Lychev, et~al.
\newblock {How Secure and Quick is QUIC? Provable Security and Performance
  Analyses}.
\newblock In {\em IEEE Symposium on Security and Privacy}, 2015.

\bibitem{mawi}
{MAWI Working Group Traffic Archive}.
\newblock \url{http://mawi.nezu.wide.ad.jp/mawi/}.

\bibitem{pir}
{Public Interest Registry}.
\newblock {Zone File Access}.
\newblock \url{http://pir.org/}.

\bibitem{fastopen2011}
S.~Radhakrishnan, et~al.
\newblock {TCP Fast Open}.
\newblock In {\em ACM CoNEXT}, 2011.

\bibitem{multipathTCP}
C.~Raiciu, et~al.
\newblock {How Hard Can It Be? Designing and Implementing a Deployable
  Multipath {TCP}}.
\newblock In {\em NSDI}, 2012.

\bibitem{draft-tls13}
E.~Rescorla.
\newblock {The Transport Layer Security (TLS) Protocol Version 1.3}.
\newblock Internet-Draft draft-ietf-tls-tls13-21, Internet Engineering Task
  Force, 2017.
\newblock WiP.

\bibitem{ietf96-quic-slides}
I.~Swett.
\newblock {QUIC - Deployment Experience @Google}.
\newblock
  \url{https://www.ietf.org/proceedings/96/slides/slides-96-quic-3.pdf}.

\bibitem{verisignTLD}
Verisign.
\newblock {Zone Files For Top-Level Domains (TLDs)}.
\newblock \url{verisign.com}.

\bibitem{verisignReport}
Verisign.
\newblock The verisign domain name industry brief.
\newblock \url{https://www.verisign.com/assets/domain-name-report-Q22017.pdf},
  Sept. 2017.

\end{thebibliography}
